# Improvement of Printing Quality for Laser-induced Forward Transfer based Laser-Assisted Bioprinting Process using a CFD-based numerical model

Jie Qu[a,b], Chaoran Dou[c], Ben Xu[d*], Jianzhi Li[c*], Zhonghao Rao[b,], Andrew Tsin[e]

a. Department of Mechanical Engineering, The University of Texas Rio Grande Valley, Edinburg, TX 78539, USA
b. School of Electrical and Power Engineering, China University of Mining and Technology, Xuzhou 221116, China
c. Department of Manufacturing and Industrial Engineering, The University of Texas Rio Grande Valley, Edinburg, TX 78539, USA
d. Department of Mechanical Engineering, Mississippi State University, Mississippi State, MS 39762, USA
e. Department of Molecular Science, The University of Texas Rio Grande Valley School of Medicine, Edinburg, TX, 78539, USA

**Corresponding author**
Ben Xu: Tel.: +1 (662) 325-5632; Email: xu@me.msstate.edu
Jianzhi Li: Tel.: +1 (956) 665-7329; Email: jianzhi.li@utrgv.edu

**Abstract:**

As one of the three-dimensional (3D) bioprinting techniques with great application potential, laser-induced-forward-transfer (LIFT) based laser assisted bioprinting (LAB) transfers the bioink through a developed jet flow, and the printing quality highly depends on the stability of jet flow regime. To understand the connection between the jet flow and printing outcomes, a Computational Fluid Dynamic (CFD) model was developed for the first time to accurately describe the jet flow regime and provide a guidance for optimal printing process planning. By adopting the printing parameters recommended by the CFD model, the printing quality was greatly improved by forming stable jet regime and organized printing patterns on the substrate, and the size of printed droplet can also be accurately predicted through a static equilibrium model. The ultimate goal of this research is to direct the LIFT-based LAB process and eventually improve the quality of bioprinting.

# INTRODUCTION

3D bioprinting is an emerging technology that has been investigated in fields varying from printing of live cells to biosensors fabrication and from stem cell fabrication to artificial organ generation (*1-3*). 3D bioprinting has gained special momentum in generation of the 3D functional tissues and organs due to its capability of periodic arrangement of various biological materials in a precisely controlled manner (*4*). As one kind of the 3D bioprinting techniques, laser assisted bioprinting (LAB) can print biological materials with as small as cell-level resolution, therefore by controlling the cell density and organization, LAB potentially holds a great promise to fabricate living tissues or organs with biomimetic physiological functionality (*5*). LAB is based on the principle of laser induced forward transfer (LIFT), which was first proposed by Bohandy *et al.* in 1986 (*6*) as an accurate solid deposition technology with high resolution. LIFT uses a pulsed laser beam focused through a transparent glass/quartz plate onto a donor layer coated on the other side of the plate to eject a tiny volume of the donor material towards a receiving substrate (*7*). The bioink transfer in LAB process is believed as the key to the formation and growth of a vapor bubble and a jet because of the rapid evaporation caused by the high energy laser pulse (*8, 9*). LIFT-based LAB has great advantages over other bioprinting technologies. These advantages include non-contact printing, high fabrication precision and high adaptability, supporting different cell patterns with good cell viability (~85%) (*2*). LIFT has similar functionality to droplet-on-demand inkjet printing (nozzle-based), however, since LIFT is a nozzle-free process, it does not suffer from nozzle clogging and compatibility issues between bioink and nozzle's materials, which provide the possibility to print bioink with a variety of properties (viscosity, and density etc.) (*10*).

Due to these advantages, LIFT-based LAB has drawn attentions from researchers and practitioners for its potential application in printing tissue or organs (*11-16*). Nevertheless, the main drawback of LIFT-based LAB is also due to its high resolution, resulting in a low flow rate, therefore it may experience some difficulty to accurately position cells on the receiving substrate (*17-19*). In addition, even though the nozzle-free feature resolves the clogging issue, it in turn has no restrictions to the flow direction and the jet regime, since the bioink transfer process completely depends on the formation of jet flow, therefore if the flow and jet regime cannot be controlled precisely, the process could suffer from deteriorated printing quality. As shown in Fig.1, when the jet flow is not fully developed, no bioink can be transferred from the coated quartz to the receiving substrate. Even if the bioink can be transferred, there are still two scenarios which may affect the printing process: the plume and the splashing cases, which actually will lead to unorganized printing pattern on the substrate with irregular droplet surrounded by many splashes. Those two printing patterns are not acceptable for precise bioprinting and the scattered droplet distribution strongly influences the final printing quality as well as the cell viability. Fig. 1 shows that only the stable jet can achieve controlled printing patterns with organized and circular droplets, therefore this is the only transfer scenario that allows for precise printing with a good printing quality and high cell viability. Consequently, a deep understanding of the jet flow regime is critical to the adoption of LIFT-based LAB process.

As agreed in a few investigations reported, a variety of printing parameters could affect the jet flow regime and in turns the printing patterns on the substrate. These parameters include the pulse laser energy intensity (*20-22*), the focal spot size (*23*), the liquid layer thickness, material properties (*5, 19*) and so on. Therefore, it is extremely difficult to theoretically model the formation of jet flow because of its nature of complex multiphysics and multiscale phenomena involved in the LIFT based LAB process. For example shock wave (*24*), plasma generation (*25*) and irradiation (*26*), are reported in the laser-liquid interaction during the LIFT-based bioprinting process. Meanwhile, the laser-liquid interaction occurs in an extremely fast manner with a typical time duration ranging from $10^{-10}$s to $10^{-12}$s, while the jet development process could take a time period ranging from $10^{-3}$s to $10^{-6}$s. These multiscale time duration will certainly complicates the attempt of developing accurate mathematical models. As a result, most reported studies required tedious experimental efforts to explore the relationship between the jet flow regime and the final printing outcomes, in order to fully understand the relationship between the process parameters and the formation of a stable jet.

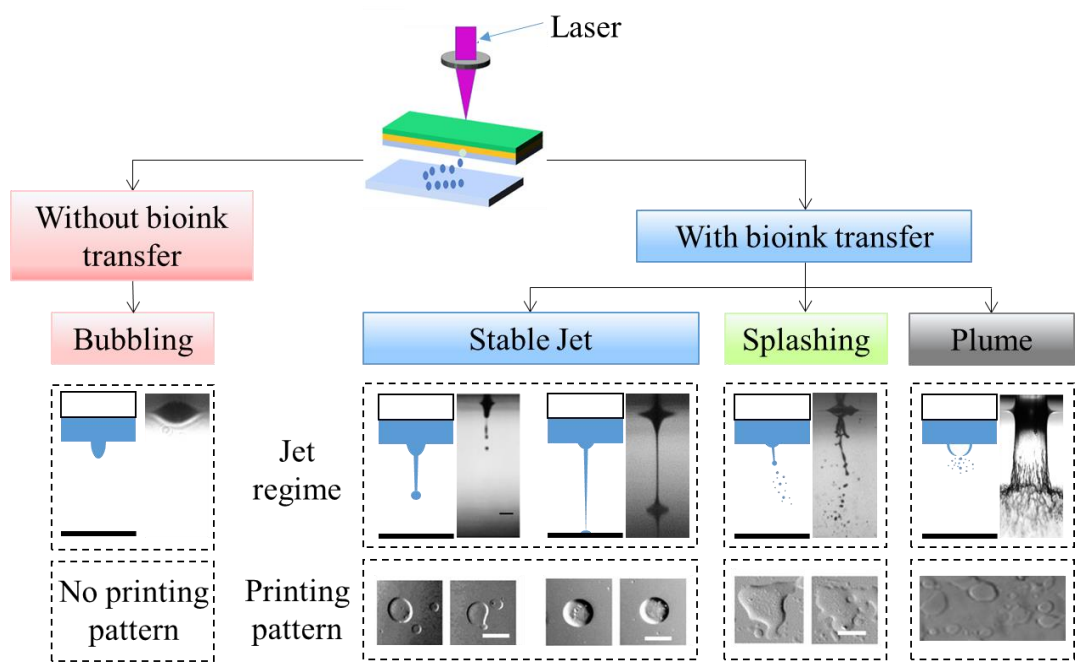

*Fig. 1. Different Jet regime and corresponding printed pattern (17, 19, 22, 27)*

Computational Fluid Dynamics (CFD) simulation is a very popular approach to predict the formation of jet and bubble in various multiphase transport processes. It could bring a good opportunity for reducing the tedious experimental efforts required in investigation of the LIFT-based LAB process. However, considering the complex multiphysics phenomena at the very beginning stage of LAB, modeling the laser-liquid interaction process from a multiscale point of view in a very concise and consistent way becomes extremely difficult. Through literature review, while there are investigations that attempted to explain thoroughly the laser-liquid interaction mechanism in LAB, most of the current work either ignored the initial bubble forming process, or relied on experimental observations by missing key information in small scales, based on which, assumptions are made. For example, Brown *et al.* (*28*) and Kalaitzis *et al.* (*29*) chose to experimentally track the interface deformation during the bioprinting, and then utilized the experimental results as the moving boundary condition to model the liquid movement and the jet. This model highly relied on the earlier experimental results, therefore it is only applicable under specific conditions, such as the same energy input, the same liquid layer thickness, and the same liquid properties. The other model is the initial bubble model (*30, 31*), which assumes that the input laser energy is converted into the internal and kinetic energy of an initial bubble. Most of the published works, which adopted the initial bubble model, chose the properties and dimensions of the initial bubble (such as the size, pressure and temperature) based on their own experiments. However, the laser energy intensity, the donor layer thickness and the position of laser focal point have strong impacts on the formation of jet and bubble (*21*), therefore it is extremely hard to extend the reported models to explore the LIFT process when these parameters are changed (*30-32*). Consequently, it is desired to develop a generalized and solid model to determine the properties and dimensions of such an initial bubble, then this generalized model can be incorporated in the CFD simulation in order to precisely model the entire LIFT based LAB process.

In the present work, a novel generalized mathematical model was developed for the first time to accurately determine the size, pressure and temperature of the initial bubble based on the energy conservation law, and then a CFD study by incorporating the proposed generalized mathematical model for the initial bubble was performed for the first time to predict the formation of jet flow and the final printing pattern on the receiving substrate. The proposed CFD-directed simulation model was validated and shown its capability of precise prediction of the jet flow behavior. Furthermore, by utilizing the simulation results as parameters input, a static equilibrium model was employed to accurately predict the size of the printed droplet. Meanwhile, a LIFT-based LAB experimental platform was built and utilized to perform more experimental works by altering the printing parameters, where a femtosecond pulse laser with 1040nm wavelength and a maximum pulse lase energy of 40μJ was adopted in this study, as shown in Fig. 2. Deionized water with dye was selected as the liquid layer for all the experimental cases. The printing quality with various printing parameters was analyzed in detail using the proposed CFD model. By adopting the printing parameters recommended by the CFD model, the printing quality was greatly improved by forming stable jet regime and organized printing patterns on the receiving substrate, and the size of printed droplet can be accurately predicted through the static equilibrium model. The ultimate goal of this research is to develop a solid connection by utilizing the proposed CFD model to direct the LAB process and improve the printing quality.

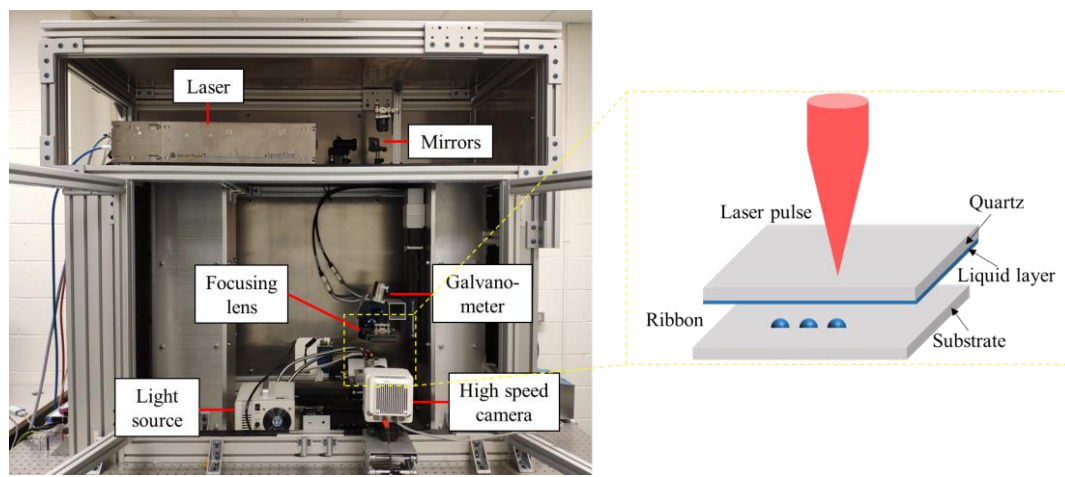

*Fig. 2. Experimental platform of LIFT process*

**RESULTS AND DISCUSSION**

In this study, we first performed LIFT bioprinting experiments with non-optimized parameters, such as the liquid layer thickness and pulse laser energy intensity. Not surprisingly, the unstable jet regime was formed so the printing quality was fairly low with unorganized printing outcomes and irregular droplets on the receiving substrate. CFD study was then performed so that the appropriate combinations of printing parameters were identified, and the bioprinting experiments were conducted one more time to verify the predicted results. Eventually, the printing quality was greatly improved by forming stable jet regime and very organized printing patterns on the receiving substrate.

**First attempt to obtain well-organized printed droplets**

As shown in Fig. 3, a laser generator (Spirit One 1040-8) was chosen to generate the pulse laser, and the laser intensity distribution satisfies the Gaussian distribution. The laser's wavelength is 1040nm, its maximum pulse energy is 40μJ, and the pulse duration is 300fs. In the experiment, every laser pulse was reflected by the mirrors and went through the galvanometer and the focusing lens, eventually focused on the ribbon, which is a quartz with liquid layer coated at the bottom. The radius of laser focal spot is 30μm, and the thickness of the quartz is 0.64cm. Deionized water was selected as the liquid layer. To enhance the absorption rate of deionized water, 1% w.t. of graphene solution was added as a dye, which can also introduce an additional benefit of biocompatibility when the actual bioink is used in the printing process. Since the liquid layer thickness was selected from 1μm to 100μm (*7*), for the first attempt in this study, an median liquid layer thickness was selected as 50μm while the pulse laser energy was varied from 10μJ to 40μJ.

Fig. 3 shows the liquid transfer and printing patterns with 50μm thick liquid layer and various pulse laser energies. It is important to note that there are two mirror lines at the top and bottom part of these figures, because the reflection of the two substrates in the figures. To clearly show the printing patterns, the mirror line at the bottom was marked by a light blue dash line. From Fig. 3A-D, the jet flow is separated as two stages: the first stage shows that a thin jet flow came out from the cone-shape structure as marked by the red dash line, and it only needed a short time period to complete the liquid transfer until 58.8μs, as shown in Fig. 3A-D; the second stage demonstrates the development of the cone-shape structure, which can be developed into two sub-stages: 1) the formation of a jet and a single droplet underneath; 2) the collapse of jet to complete the liquid transfer. The second stage took a longer time to complete than the first stage. For the case with pulse laser energy of 10μJ or under, the first stage needed about 176.4μs to be completed (Fig. 3A). However, after 176.4μs, since the pulse laser energy was too small to develop the cone-shape well, the second stage liquid transfer process could not be completed. Once the jet collapsed at 176.4μs, the droplet started to move upward instead of downward. With the development of jet flow, the droplet underneath the ribbon tended to move downward due to the remaining momentum from the bubble's fast expansion, while the cone-shape structure had a tendency to move upward and bounce back to the liquid layer below the ribbon due to the collapse of bubble and the surface tension, and provided a force pointing upward. The opposite movement direction between the droplet and the cone-shape structure eventually led to the break of jet, as shown in Fig. 3A at 235.2μs. For the case with pulse laser energy of 10μJ, the upward momentum from the cone-shape structure was dominated over other effects, therefore it made the droplet bounce back to the liquid layer. In this case, the printed droplet with 10μJ laser energy has the smallest diameter, as shown in Fig. 3E. For jet flow with laser pulse energy of 20μJ and 30μJ, the first stage can be completed before 117.6μs with stable jet regime (Fig. 3B and C). And the second stage was well developed with the droplet nearly touching the substrate. With a bigger energy input, the remaining downward momentum was dominated, so the underneath droplet moved downward after jet broke, therefore the second liquid transfer stage can also be completed. The major difference between the jet flow with 20μJ and 30μJ pulse laser energy was that the second jet was thicker and more liquid was transferred if 30μJ pulse laser energy was adopted. Apparently, both cases can print reasonably round shape droplets, and the diameters were around 187.5μm and 237.5μm, respectively. And the case with 30μJ laser energy input has a bigger droplet area.

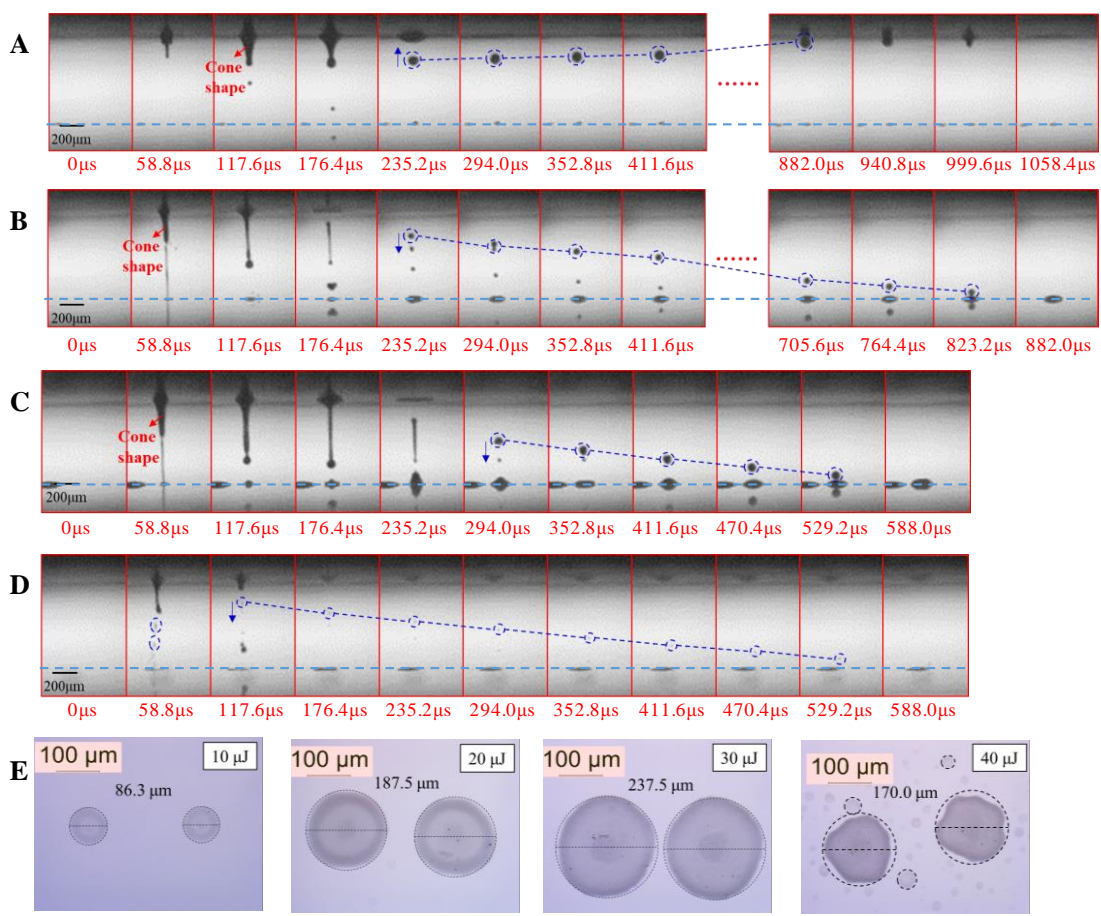

*Fig. 3. Liquid transfer and printing patterns with 50μm thick liquid layer. (A) Jet flow with 10μJ pulse laser energy. (B) Jet flow with 20μJ pulse laser energy. (C) Jet flow with 30μJ pulse laser energy. (D) Jet flow with 40μJ pulse laser energy. (E) Printing patterns of 50μm thickness liquid layer with different pulse laser energies.*

However, once the pulse laser energy was further increased, although the two stages of liquid transfer can be completed, the jet still cannot hold a stable cone shape and was turned into a splashing regime, as shown in Fig. 3D. Both the first and second jets broke into multiple tiny droplets and then scattered. The laser energy input was too big for the liquid layer to hold and develop a stable jet flow. Meanwhile, the printed droplet on the substrate with 40μJ showed a chaotic printing pattern, such that a biggest droplet was surrounded by multiple satellite droplets, as shown in Fig. 3E. The diameter of the biggest droplet was around 43.1μm, while the smallest droplet had only about 2.5μm diameter. Apparently, this type of printing pattern was not acceptable for precise LIFT-based bioprinting, because it would completely ruin the structure of printed tissue or organ.

In summary, from our first attempt of 3D bioprinting using water as the liquid, we can conclude that only a stable jet could result in well-printed outcomes, and the jet regime can predict the printing pattern based on the input laser energy. Nevertheless, a quantitative analysis cannot be developed with such limited information about the jet formation and jet regime, therefore in the next section we will discuss about the proposed CFD model and simulations.

**Numerical simulation of the development of bubble/jet flow during LIFT process**

Since the development of bubble/jet flow in the first stage occurs in a wide span of spatial and temporal scales, it is extremely difficult to monitor the printing process and tune the printing parameters in order to improve the printing quality. In addition, the first stage demonstrates most of the underlined features for the entire LIFT based LAB process, such as bubble growth and jet breakage, therefore if the development of bubble/jet flow in the first stage can be controlled precisely, the printing quality will be significantly improved and obtain well-organized printing patterns. CFD simulation is a powerful and efficient tool which can assist the design process by reducing the tedious experimental efforts. By combining CFD and the bioprinting experiment, CFD can predict the unique features of jet and bubble formation in the first stage, and direct the bioprinting process for better printing quality by recommending reasonable printing parameters based on the relationship between the jet regime and the printing patterns on the substrate.

Because the development of bubble/jet flow simulation is a multiphase process, the Volume of Fluid (VOF) model was employed to track the liquid-gas interface. The geometry and meshing configuration of the CFD model are shown in Fig. 4A. The computational domain is part of the LIFT ribbon with various thickness, 800μm width liquid layer and 900μm air in length. Only half of the model was meshed and simulated because of the axisymmetric geometry. A structured mesh was used in this case, and the mesh near the boundary was refined. The boundary conditions are also shown in Fig. 4A. The right side of the liquid layer was defined as pressure-inlet while that of air zone was defined as pressure-outlet. Besides the axisymmetric boundary condition at the axis, other boundaries were all defined as "wall". The parameters of initial bubbles were patched before simulation started. The vapor was set as the ideal gas while the liquid and air were assumed incompressible fluid. Considering the jet flow regime observed in the experiment, the laminar model was selected in the simulation. To validate the current model with the published experimental works, one case with 100μm thick 65%-glycerol layer and 717 mJ/cm$^2$ laser fluence was simulated and compared with the experimental results from literature (*17*), as shown in Fig. 4D.

Fig. 4B shows the simulation results of LIFT process with 100μm 65%-glycerol layer and 717 mJ/cm² laser fluence. It clearly demonstrated the entire development of jet flow, including generation and breakage. Firstly, once the high-energy laser pulse hit the liquid layer, the rapid evaporation of liquid generated a high pressure and high temperature initial vapor bubble. Due to the high pressure vapor inside the bubble, the initial bubble expanded rapidly. Because the quartz can be considered as a rigid wall boundary condition, the initial bubble expanded asymmetrically to a cone shape. With the bubble expansion, the high pressure inside the bubble was released and decreased. Once the pressure inside the bubble became lower than the outside atmosphere pressure, the bubble began to collapse. At this time, liquid around the tip of the bubble moved downward due to the remaining momentum from the fast bubble expansion, and then the jet flow was formed at the tip of the bubble. Meanwhile, because the viscous forces and surface tension, a reversed jet inside the bubble was also generated (*23*), as shown in Fig. 4C. With the development of both jets, the reversed jet reached a much higher velocity than that of the primary jet, for instance, the velocity of the primary jet was about 49m/s, while that of the reversed jet was 87m/s. This phenomenon was because the reversed jet was much smaller than the primary jet, and the pressure inside the bubble was lower than the outside ambient pressure.

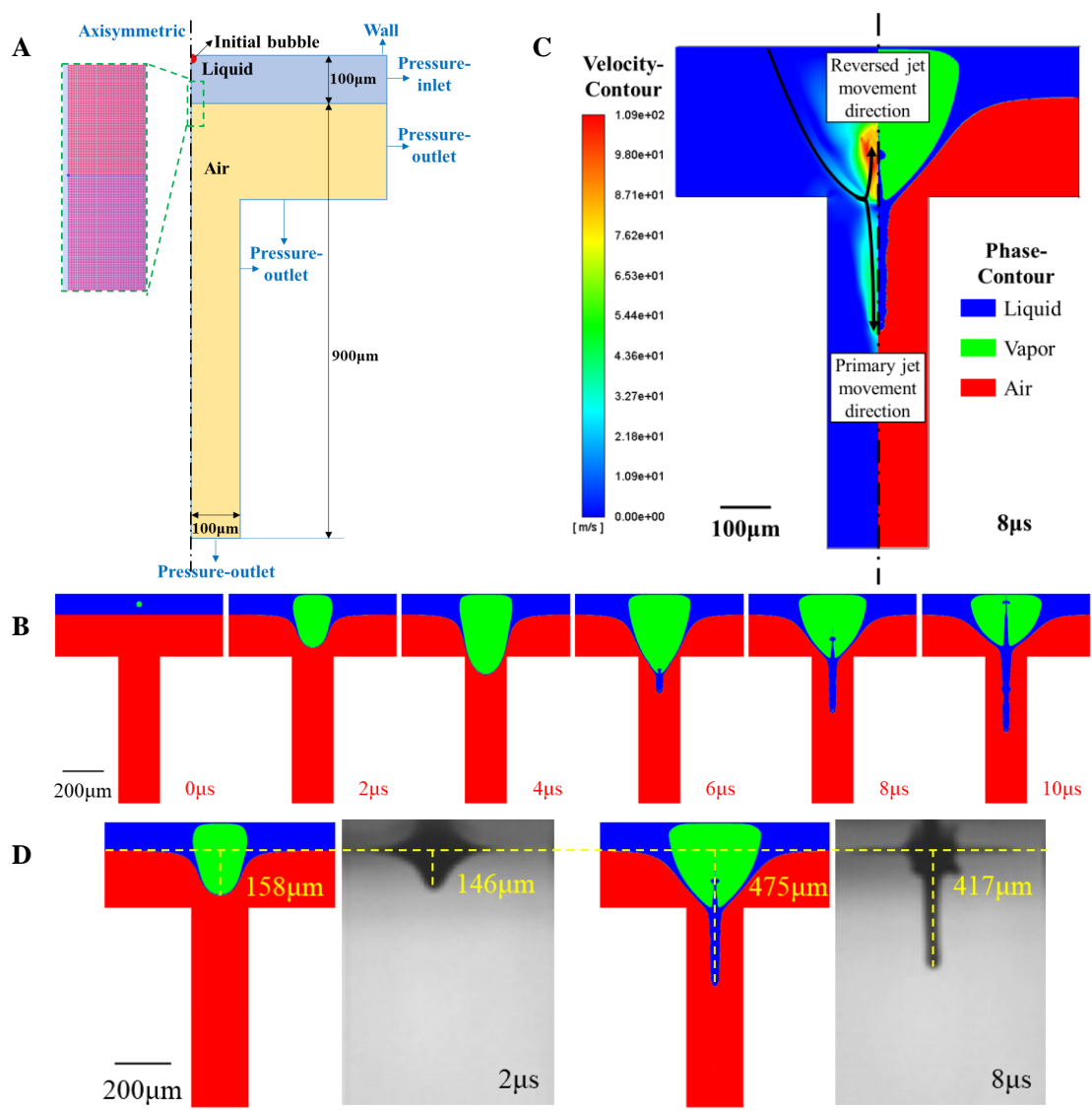

*Fig. 4. Simulation of jet flow. (A) Simulation model geometry and meshing configuration. (B) Jet flow with 100μm thickness 65% glycerol layer and 717 mJ/cm² laser fluence. (C) Velocity of jet flow at 8μs. (D) Comparison with experimental results (17).*

A comparison between the simulations and experimental results was also provided in Fig. 4D, where all the experimental conditions were maintained the same as the simulation. The length of jet in the simulation was slightly longer than that of the experiment, and the relative difference between the simulation and experiment was around 14%. Considering the associated numerical error, the proposed CFD model can be validated in a reasonable range, therefore it is trustworthy for other studies in order to identify the appropriate printing parameters for good printing quality.

Since the experimental results already showed that the liquid transfer and printing pattern were unacceptable for 50μm thick liquid layer and 40μJ pulse laser input, cases with different liquid layer thickness (50μm, 100μm, 150μm) with pulse laser energy of 40μJ were studied in this section to obtain an optimized layer thickness. Meanwhile, cases with 100μm thick liquid layer and various pulse laser input (10μJ, 20μJ, 30μJ and 40μJ) were also simulated to study the effect of pulse laser energy. Once all the simulations were completed, in the next section experiments were carried out by adopting the recommended printing parameters from the simulations. The printing parameters used in simulations and experiments are shown in Table.1.

*Table 1. Parameters for simulations and experiments*

|  | Pulse laser energy-10μJ | Pulse laser energy-20μJ | Pulse laser energy-30μJ | Pulse laser energy-40μJ |
|---|---|---|---|---|
| Liquid layer thickness-50μm | E-1 | E-2 | E-3 | E-4 / S-1 |
| Liquid layer thickness-100μm | E-5 / S-2 | E-6 / S-3 | E-7 / S-4 | E-8 / S-5 |
| Liquid layer thickness-150μm | N/A | N/A | N/A | E-9 / S-6 |

*Note: E: experiment; S-simulation.*

The simulation results of LIFT process with 50μm liquid layer and 40μJ (S-1) are shown in Fig. 5A. As discussed before, the initial bubble expanded rapidly at first. However, the liquid layer could not hold the rapid bubble expansion and therefore it was broken at about 0.5μs. Apparently the stable jet could not be formed for this case, therefore it showed a good agreement with the experiment in Fig. 3C. With the breakage of the bubble, the high pressure and high temperature vapor inside were released and then mixed with the ambient. With the same pulse laser energy input, increasing the liquid layer thickness would help to generate a stable jet. As shown in Fig. 5B and C, when the liquid layer thickness were increased from 50μm to 100μm (S-5) and 150μm (S-6), the bubble was broken first, and then the bubble kept developing and formed a regular jet flow. Because a thicker liquid layer was more capable of holding the vapor bubble, therefore it had a more robust bubble development. It is noteworthy that the length of jet for S-5 was always longer than that of S-6 at the same instant, as shown in Fig. 5D. The reason for this phenomenon is that a thicker liquid layer could provide a bigger flow resistance to slow down the rapid bubble expansion with the same laser energy input. For both S-5 and S-6, it showed a linear relationship between the length of the jet and the time duration. The maximum velocity of jet flow with different liquid layer are shown in Fig. 5E.

For S-5, the maximum jet flow velocity could reach 157 m/s at 1μs, while the maximum jet flow velocity was 89.4 m/s for S-6. In addition, with the bubble expansion, the maximum velocity decreased until the tip of the jet flow was generated. After that, the velocity was increased slightly at 4μs for S-5 and at 6μs for S-6, respectively. This phenomenon is because the liquid tip was less affected by the surface tension when it bulged out from the liquid film, therefore it kept developing to a longer jet. Furthermore, the maximum velocity eventually became stable for each case, for example, the maximum velocity of S-5 was about 110m/s, while it was about 60m/s for S-6.

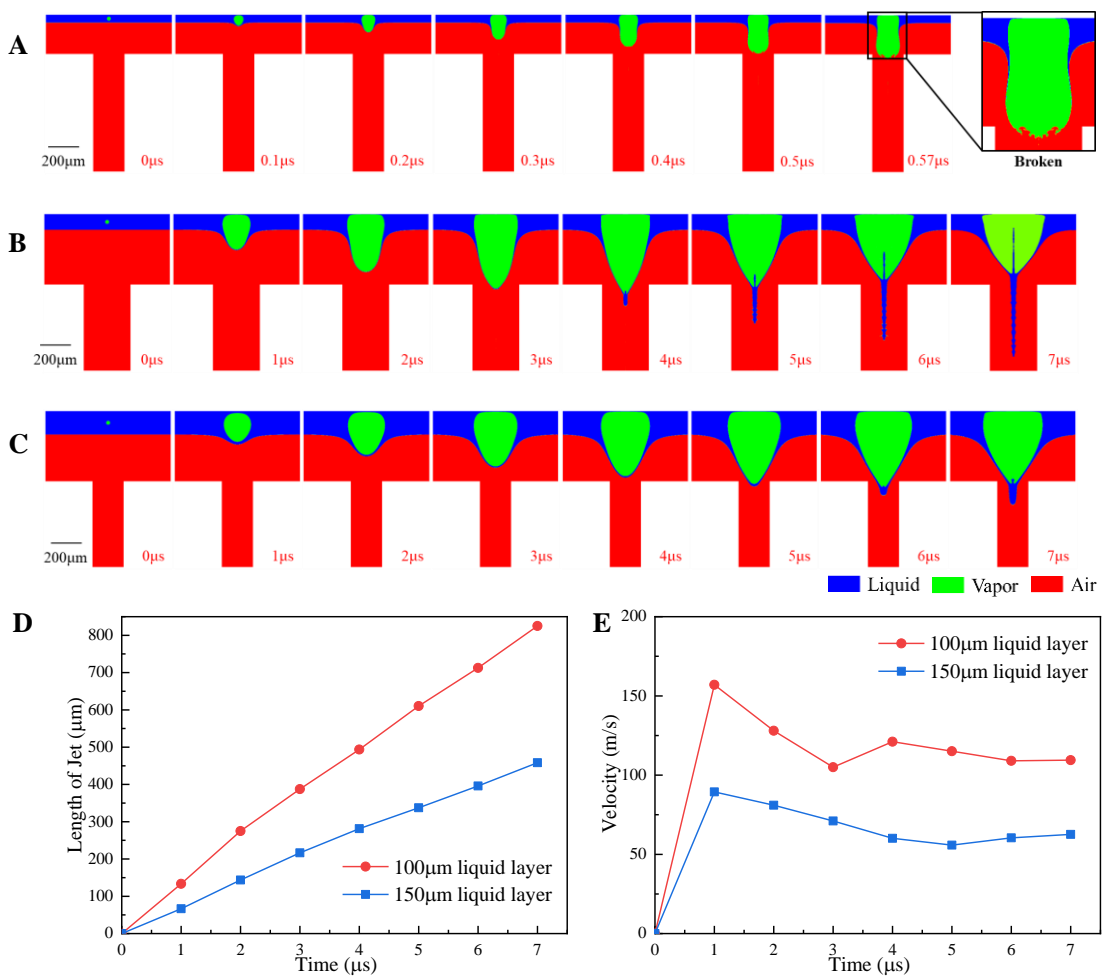

*Fig. 5. Simulation of jet flow. (A) Jet flow with 50μm thick liquid layer and 40μJ pulse laser energy. (B) Jet flow with 100μm thick liquid layer and 40μJ pulse laser energy. (C) Jet flow with 150μm thick liquid layer and 40μJ pulse laser energy. (D) The length of jet flow with different liquid layer thickness. (E) The maximum velocity of jet flow with different liquid layer thickness.*

The simulation results of 100μm liquid layer with various laser energy inputs are shown in Fig. 6. The bubble expansion and jet formation process of S-5 was already discussed, and these processes for other cases with smaller laser energy input were all very similar. Nevertheless, the size of bubble, the length of jet and the velocity were different for those cases. At the same time instant of simulation, the size of expanded bubble increased with the increasing of pulse laser energy, and the length of jet flow increased with the increasing of pulse laser energy as well. For certain pulse laser energy input, the length of the jet flow and its time duration showed a linear relationship, but the relationship between the length of the jet flow at the same instant and pulse laser energy was nonlinear (Fig. 6D). The velocity of the jet flow also increased with the increasing of pulse laser energy. With the developing of jet flow, the velocity remained almost as a constant after 4μs. The velocity of the stable jet flow with 10μJ (S-2) was about 25m/s, while it was around 70m/s for S-3 with 20μJ, increased about 180%. However, for the pulse energy changing from 20μJ (S-3) to 30μJ (S-4), the velocity was only increased 33.3%, which also showed a nonlinear relationship between the velocity and the laser energy input. Fig. 6G shows the mass flow rate versus time for cases with different laser energies. The mass flow rate was defined by the amount of liquid moved downward through the initial liquid-air interface per unit time. From Fig. 6G, the mass flow rate decreased with the development of the jet flow. Even though the tip of jet flow remained at a similar level of velocity, the whole jet flow was slowed down by the bubble collapse, and the adhesion force also provided flow resistances. In addition, it also showed a nonlinear relationship between the mass flow rate and the pulse laser energy (Fig. 6H).

To summarize this section, by adopting the proposed CFD model, different cases with various liquid layer thicknesses and laser energies were investigated numerically, and the developing of the bubble expansion and jet flow was clearly described. With the increase of liquid layer thickness from 50μm to 100μm for cases with 40μJ pulse laser energy, the jet regime developed from unstable jet to stable jet regime. With the increase of the pulse laser energy for the cases with 100μm thick liquid layer, the length and velocity of the jet both got increased. Based on the simulation results, a stable jet can be obtained by choosing 100μm liquid layer with various pulse laser energy from 10μJ to 40μJ. In conclusion, for pulse laser energy varying from 10μJ to 40μJ, the CFD simulations recommended a liquid layer thickness around 100μm for a better printing quality.

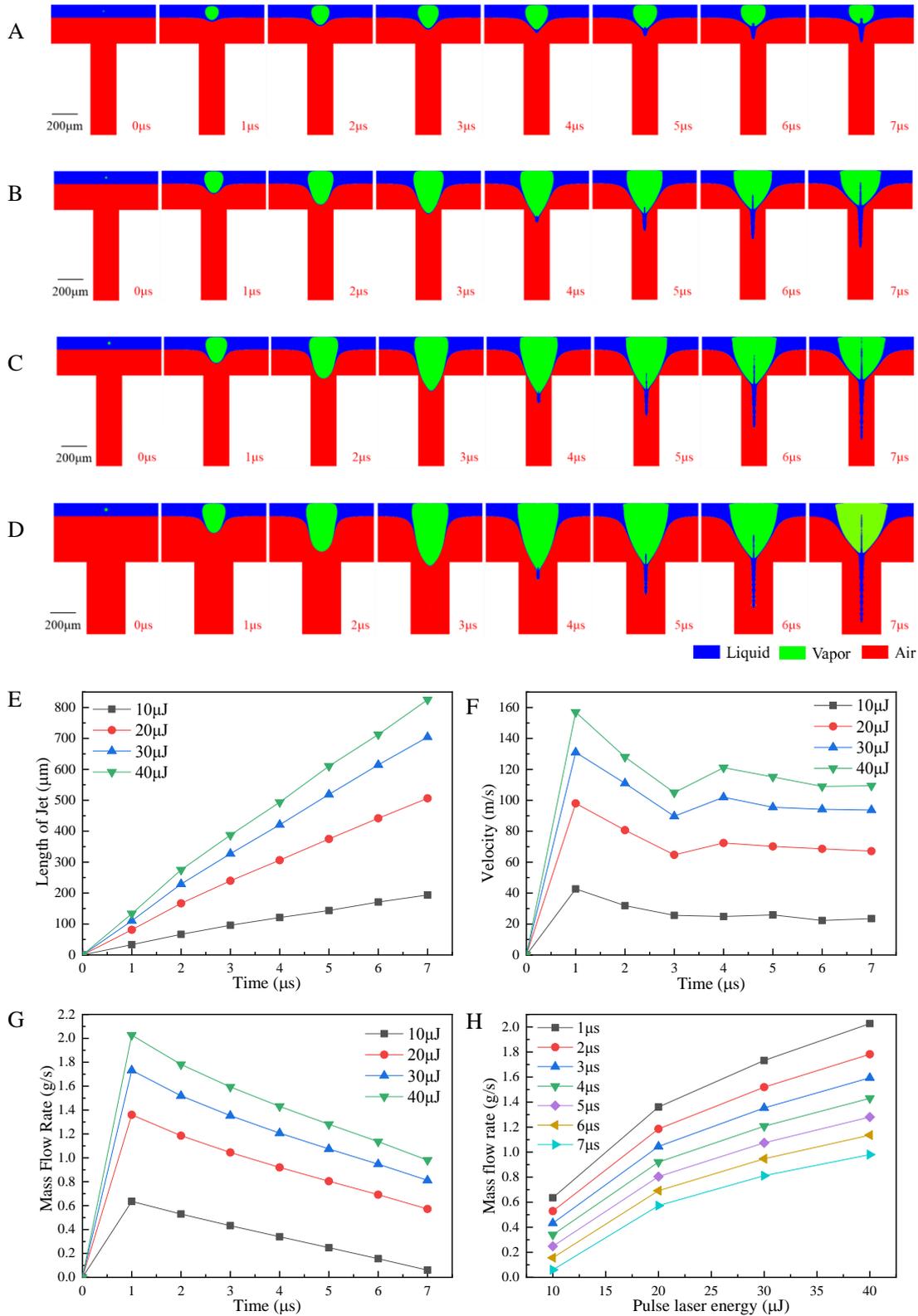

*Fig. 6. Simulation results of jet flow with different laser energy. (A) Jet flow with 100μm thickness liquid layer and 10μJ pulse laser energy. (B) Jet flow with 100μm thickness liquid layer and 20μJ pulse laser energy. (C) Jet flow with 100μm thickness liquid layer and 30μJ pulse laser energy. (D) Jet flow with 100μm thickness liquid layer and 40μJ pulse laser energy. (E) The length of jet flow with different laser energy. (F) The maximum velocity of jet flow with*

*different laser energy. (G) The mass flow rate of jet flow versus time with different laser energy. (G) The mass flow rate of jet flow versus laser energy.*

**Printed droplets after optimization**

In this section, we tried to utilize these recommended printing parameters to experimentally print out the droplets and also find out the connection between the size of printing pattern and the characteristics of jet flow.

The liquid transfer of 150μm and 100μm thick water layer with 40μJ pulse laser energy are shown in Fig. 7A and B. No jet flow and liquid transfer were observed when the liquid layer thickness was 150μm (Fig. 7A). Compared with 50μm liquid layer (Fig.3), the same amount of pulse laser energy input could not provide adequate pressure to overcome a bigger flow resistance. The generated bubble still could be expanded, but it only formed a peak at 117.6μs, and started to collapse afterwards. At about 400μs, the liquid layer returned to a flat surface at the upper layer. When the liquid layer thickness was 100μm (Fig. 7B), a complete process of both the first and second stages of jet flow was formed. The jet flow in the first stage was connected with the substrate at 58.8μs while in the second stage it was connected with the substrate at about 235.2μs. This happened because the laser energy input was large enough to drive the jet flow and develop to a sufficient length, at the same time the jet with the 100μm thick liquid layer (E-8 in Table 1) was also robust enough not to break during the jet developing. Later on the linkage between the liquid layer and the substrate became thinner and thinner, and eventually detached from the top liquid layer between 411.6μs and 470.4μs, as shown in Fig. 7B. The broken linkage finally formed a droplet due to the surface tension and fell on the substrate by completing the second stage of liquid transfer.

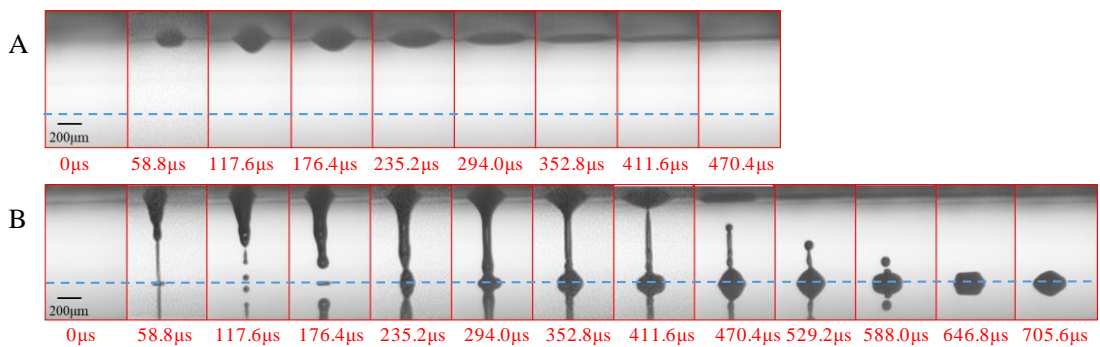

*Fig. 7. Liquid transfer with different thickness liquid layer and pulse laser energy. (A) Jet flow of 150μm thick liquid layer with 40μJ pulse laser energy. (B) Jet flow of 100μm thick liquid layer with 40μJ pulse laser energy.*

More cases with 100μm thick liquid layer and different pulse laser energy inputs were experimentally investigated, and the liquid transfer and printing patterns were shown in Fig.8. As predicted by the CFD studies in the previous section, a stable jet can be formed for the cases with 100μm thick liquid layer and pulse laser energy varying from 10μJ to 40μJ. The test results actually demonstrated that the jet flow process with 100μm thick liquid layer showed very similar phenomenon as the case with 50μm thick liquid layer, as shown in Fig. 8 A-C, where the connection between the jet and the liquid layer became much thinner while maintaining the liquid transfer, and a separated droplet (marked by the blue dash circle) was formed on top of the primary droplet. The gourd-shaped droplet was also formed and can be observed at 235.2μs in E-5, 294.0μs in E-6 and E-7. However, the gourd-shaped droplet could not be detected when the liquid layer thickness was 50μm. Therefore, the jet flow remained more robust for the cases with thicker liquid layer than other cases, for instant the liquid layer thickness of jet in E-2 and E-6 were 33.8μm and 92.9μm at 176.4μs, respectively.

Fig.8 A-C shows the moving trajectory of the separated droplet, which was marked by the blue dash circle. For the liquid transfer process with 50μm thick liquid layer, the velocity of the separated droplet increased with the increasing of pulse laser energy. When the pulse laser energy reached to 40μJ, the jet regime changed from the stable jet to the splashing jet mode, but the velocity of the separated droplet was not affected too much by the pulse laser energy input for the case with 100μm thick liquid layer (Fig.8 F), it was probably because the separated droplet was almost static for the stable jet, and the initial velocity of the separated droplet was almost zero. With the assistance of gravity, the separated droplet then fell onto the receiving substrate. With such a short distance between two substrates and such a short time period, the falling velocity was about the same for all the cases with different pulse laser energy inputs. However, when the pulse laser energy reached to 40μJ (Fig.8 D), the jet flow could directly connect with the substrate, and no separated droplets were formed. Similarly, the size of printed droplet on the substrate increased with the increase of pulse laser energy. As shown in Fig.8 G, for the same pulse laser energy, the size of printed droplet on the substrate for the case with 100μm thick liquid layer was bigger than the case with 50μm thick liquid layer. Considering the unstable jet regime of liquid transfer process with 50μm liquid layer thickness and pulse laser energy of 10μJ and 40μJ, the droplet size of these cases was not typical. For case with 100μm thick liquid layer, it showed a linear relationship between the droplet size and the pulse laser energy, which confirmed the conclusions from Lin *et al.* (*8*) and Kattamis *et al.* (*33*), because their results also indicated that both the LIFT process with/without an absorption layer showed a linear relationship between the droplet size and the laser energy input, therefore they shared a similar mechanism of liquid transfer.

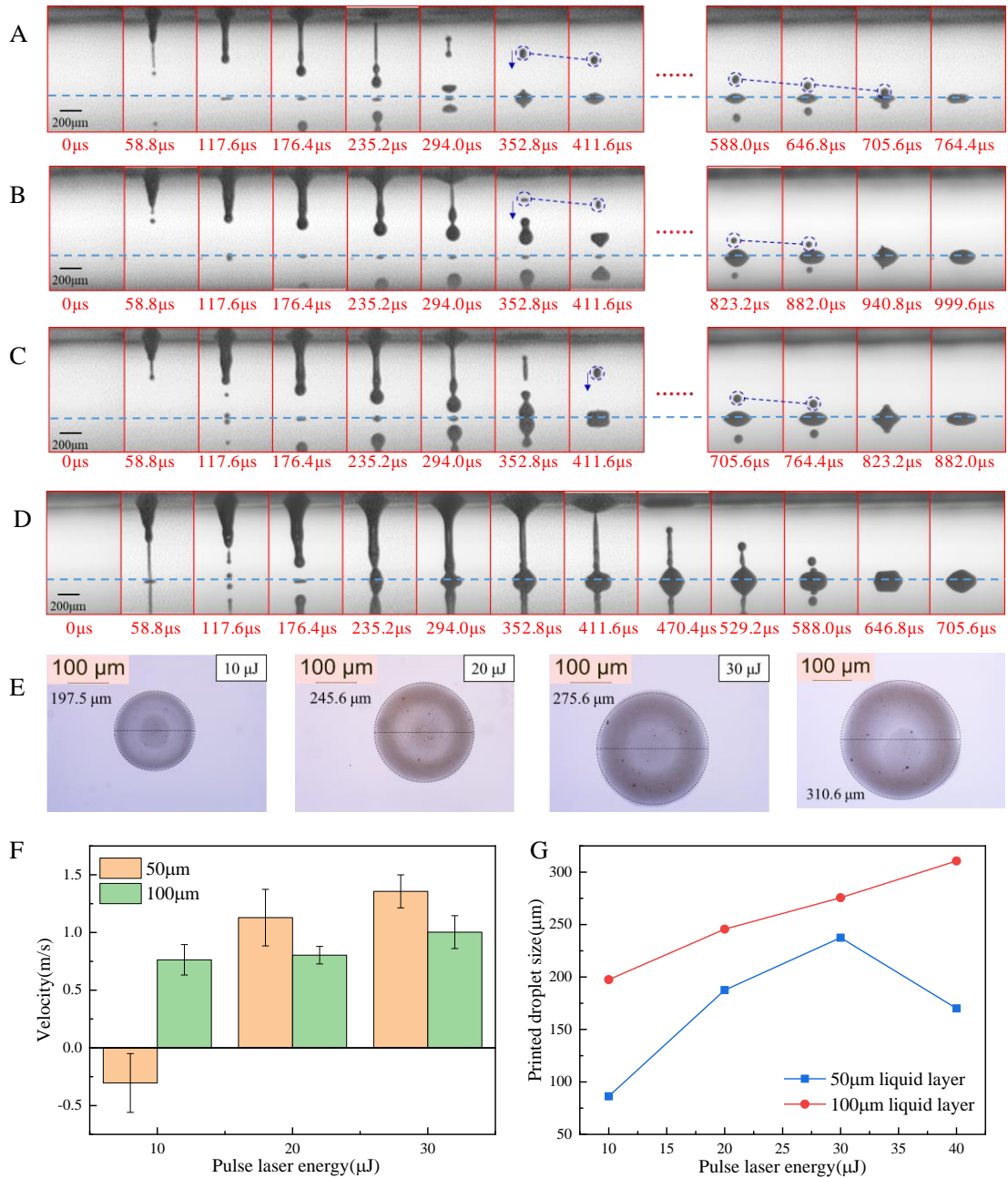

*Fig. 8. Liquid transfer and printing patterns with 100μm thick liquid layer. (A) Jet flow with 10μJ pulse laser energy. (B) Jet flow with 20μJ pulse laser energy. (C) Jet flow with 30μJ pulse laser energy. (D) Jet flow with 40μJ pulse laser energy. (E) Printing patterns with different pulse laser energies. (F) The movement velocity of the dropped droplet. (G) Printed droplet size of different liquid layer thickness with different pulse laser energy.*

Based on the discussions above, the printing parameters recommended by the CFD simulation were proved to ensure a stable jet regime and improve the printing quality. Because the initial jet flow significantly affects the printing quality and the size of printed patterns on the substrate, a quantitative analysis is desired to reveal the relationship between the jet flow and the size of the printing pattern. A regression curve fitting and a static equilibrium model were developed in this study to predict the size of printed droplet by utilizing the simulation results as input parameters, and the experimental results were utilized to verify the prediction. The flow chart of comparison strategy between simulation results and experimental results is shown in Fig. 9.

Based on the conclusion from van Dam & Le Clerc (*34*), the velocity and volume of droplet are the two main factors that influence the size of printing pattern. Since we already got the moving velocity of the jet flow from the simulations, we can show the transferred liquid volume versus the mass flow rate obtained from the simulation, as shown in Fig.10 A. Apparently, the volume of the transferred liquid and the mass flow rate showed a linear relationship, and a regression model can be obtained as $V = 1.13 \times 10^{-6} \times \dot{m}/\rho$. The coefficient of this curve fitting equation is $1.13 \times 10^{-6}$, which is related to the developing time of jet flow and the distance between two substrates.

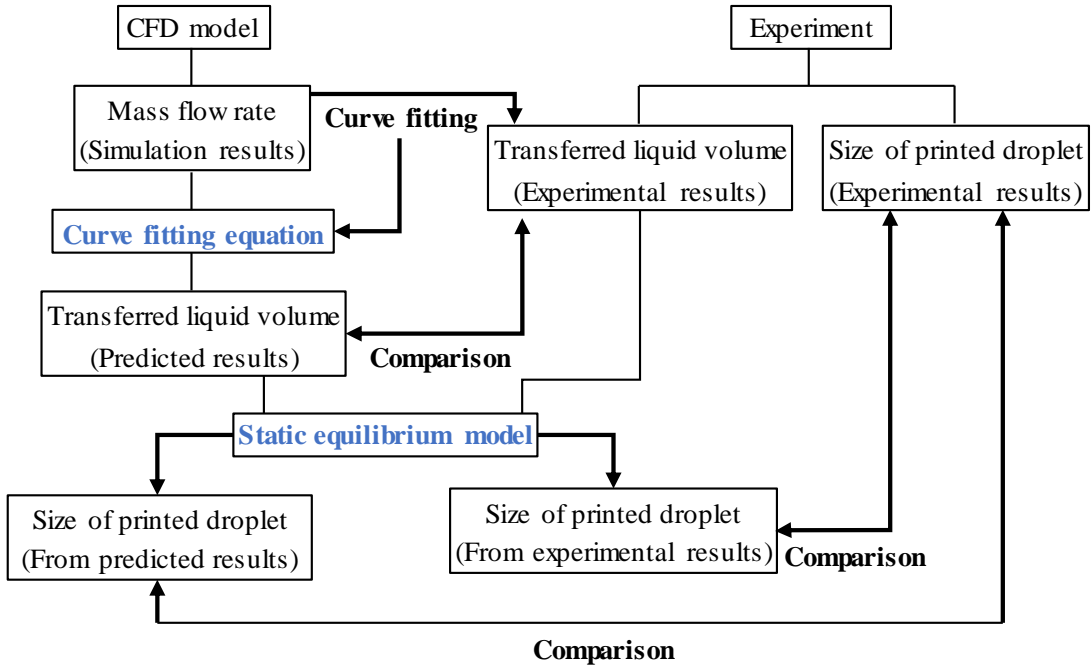

*Fig. 9. Flow chart of comparison strategy between simulation results and experimental results*

In addition, we can also utilize a mathematical model to predict the maximum size of printed droplet on the receiving substrate, and then compared with the experiment, as shown in Fig.9. Since the droplet on the substrate is in the static state, the size of the droplet was only related to the volume, surface tension of liquid and the surface properties of the substrate. Assuming the droplet as part of sphere shape, the static equilibrium equation was shown as follows (*35*),

$$-2\pi\sigma(l - r\cos\theta) + 2\pi\sigma\left[R\arcsin\left(\frac{r}{R}\right) + \frac{r(z^2 - r^2)}{z^2 + r^2}\right] + \rho g \pi\left[R^3 \arcsin\left(\frac{r}{R}\right) - \frac{1}{4}\frac{r^5}{z^2} + \frac{1}{4}z^2 r - \frac{2}{3}r^3\right] = 0 \quad (1)$$

where $\sigma$ is the surface tension, $\theta$ is the contact angle, $\rho$ is the density, $r$ is the radius of the droplet, $z$ is the height of the droplet, $R$ is the radius of the sphere, $l$ is the arc length of the droplet. $z$, $R$ and $l$ could be defined as below:

$$z = \left[\frac{3V}{\pi} + \left(r^6 + \left(\frac{3V}{\pi}\right)^2\right)^{1/2}\right]^{1/3} - r^2\left[\frac{3V}{\pi} + \left(r^6 + \left(\frac{3V}{\pi}\right)^2\right)^{1/2}\right]^{-1/3} \quad (2)$$

$$R = \frac{r^2 + z^2}{2z} \quad (3)$$

$$l = R\arcsin\left(\frac{r}{R}\right) \quad (4)$$

where $V$ is the volume of the droplet. Based on the discussions above, the volume could be obtained from both the experimental results or the predicted results calculated by the curve fitting equation.

Eq. (1) was adopted to calculate the size of the printed droplet by using the transferred liquid volume from experimental results and predicted results, and the comparison between the experiment and simulation is shown in Fig.10 B. Both the simulation and experimental results showed that the sized of printed droplet increased with the increasing of pulse laser energy. Meanwhile, both the size of the printed droplet calculated from transferred liquid volume of experimental results and the predicted results showed good agreement with the actual measured droplet size, while the simulation results with liquid volume from curve-fitting prediction as input were closer, especially with pulse laser energy of 40μJ. Utilizing this static equilibrium model can directly connect the size of printed droplet with the simulation results, as shown in Fig. 10C. Furthermore, the static equilibrium model can also be combined with the proposed CFD model to predict the jet flow regime and the size of printed droplet, and it can provide a great guideline to direct the design of experimental process.

Eventually, a well-organized printed pattern on the receiving substrate with different alphabets is shown in Fig.10 D, where UT means University of Texas, and CUMT is the abbreviation of China University of Mining and Technology, and this is a successful demonstration of CFD-based improvement of printing quality for LIFT-based LAB process.

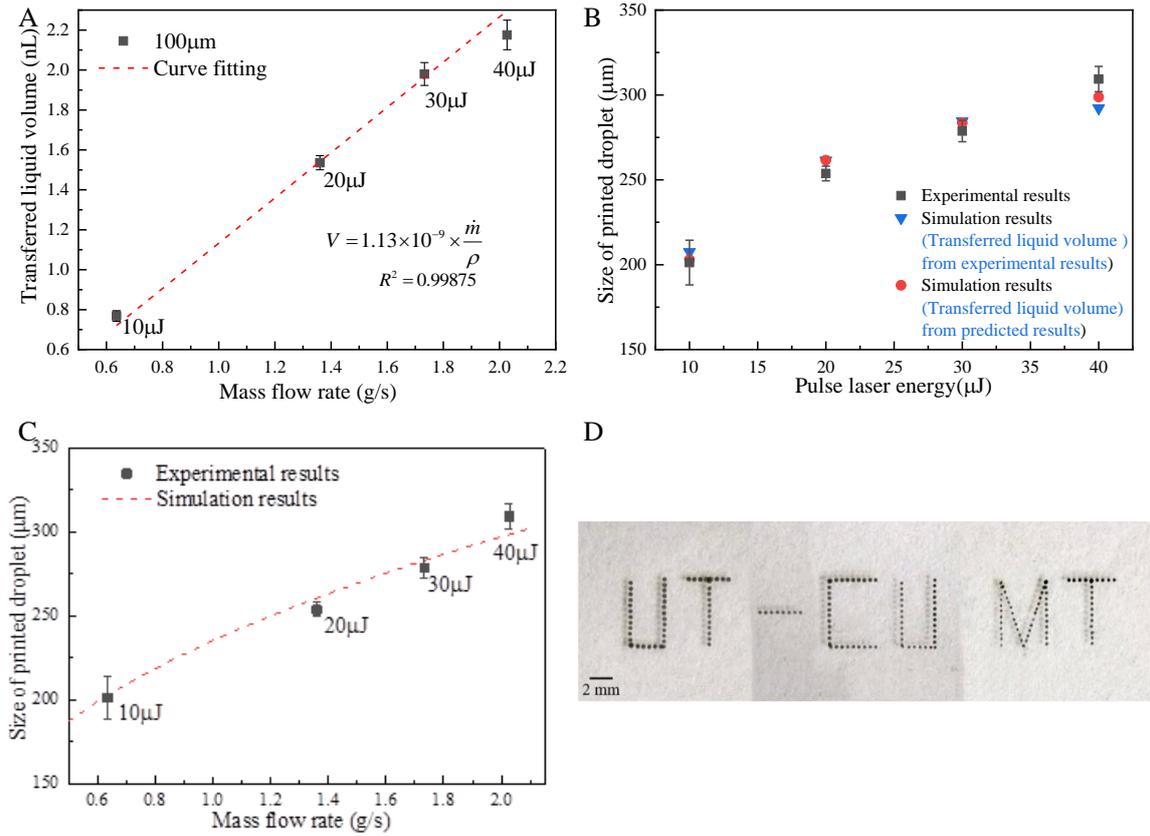

*Fig. 10. (A) Prediction of transferred liquid volume with the mass flow rate from simulation results. (B) Experimental and simulation results of printed pattern size. (C) Experimental and simulation results of printed pattern size versus mass flow rate. (D) Printed pattern after optimization (UT: University of Texas, CUMT: China University of Mining and Technology)*

## CONCLUSIONS

The major contribution of this work is to develop a CFD model to guide the LIFT-based LAB for the first time in the bioprinting research community, and this model provides a great opportunity to quantitatively predict the generation and development of bubble and jet flow in the LIFT-based LAB process, and eventually improve the final printing quality by adopting the appropriate printing parameters recommended by the CFD model. The numerical model was validated by the experimental results, and a good agreement was achieved in terms of the size of printed droplet. By utilizing the proposed CFD model, this study demonstrated a successful example of well printed pattern, as shown in Fig. 10D. The key conclusions are listed as follows:

(1) The liquid layer thickness strongly affects the formation and development of jet flow. A thin liquid layer cannot maintain the jet flow due to the rapid bubble expansion with large pulse laser energy input, therefore the jet eventually would break and reach to the splashing jet regime. Furthermore, the jet cannot be formed when the liquid layer was too thick.

(2) For all the stable jets investigated in this study, the length of the jet flow and the time duration of jet flow showed a linear relationship, as shown in Fig. 5D. With the development of jet, the velocity of the jet flow remained almost as a constant. A reversed jet inside the bubble was also observed because of the viscous forces and surface tension.

(3) For cases with the same liquid layer thickness, the size of the printed droplet, the

velocity and length of the jet flow all increased with the increase of pulse laser energy, as shown in Fig. 6 and Fig.8 G.

(4) Utilizing the simulation results, the volume of transferred liquid trough LIFT-based LAB process could be accurately predicted. With the assistance of static equilibrium model describing the static balance of droplet and substrate, the size of printed droplet can also be predicted, as shown in Fig.10 B and C.

## MATERIALS AND METHODS

### Description of experiments

The experimental platform can be found in Fig. 1. An XY stage (Pro115LM Aerotech) was utilized to move the substrate up and down to get different print patterns. A light source (HL150-A Fisher Scientific) was used to provide a sharp background, and a high-speed camera (Phantom VEO 410L) was adopted to monitor and record the LIFT printing process. Several high magnification zoom lenses (Navitar) were utilized to obtain videos and images with high resolutions. The frame rate was set as 57,000 fps and the exposure time was fixed as 3μs. In addition, a microscope (LEICA MC 170 HD) was utilized to observe and record the printed droplet patterns on the substrate for more analysis.

### Modeling - Initial bubble parameters

As shown in Fig. S1, the laser energy distribution $E$ in this study was adopted as a Gaussian distribution (*25, 36, 37*),

$$E(r) = \frac{E_0}{\sigma\sqrt{2\pi}} \exp\left(-\frac{r^2}{2\sigma^2}\right) \tag{1}$$

where $E$ is the pulse energy at different position, $E_0$ is approximated as a 99.7% distribution range, $r$ is the position of interest, $\sigma$ is the spatial standard deviation of laser beam profile.

Due to the Gaussian distribution of laser energy, the energy increases toward the center while decreases toward the edge. Assuming a threshold of laser fluence exists to define the laser interaction diameter, while only the liquid layer inside this interaction area could absorb the laser energy input for phase change and temperature increase. The threshold can be defined by utilizing the energy input at $r_T$ divide the area of ring around $r_T$,

$$F_T = \frac{E(r_T)}{\pi(r_T + \Delta r)^2 - \pi(r_T - \Delta r)^2} \tag{2}$$

where $F_T$ is the threshold of laser interaction fluence, $r_T$ is the laser interaction radius, $\Delta x$ is the half of width of the ring near the laser interaction radius. For different types of lasers and liquids, the threshold of laser interaction fluence should be different.

After calculating the laser interaction radius $r_T$, the energy absorbed by the liquid layer can be calculated by integrating the laser energy distribution from $-r_T$ to $r_T$,

$$E_a = \int_{-r_T}^{r_T} E(r)dr \tag{3}$$

where $E_a$ is the absorbed energy by the liquid layer.

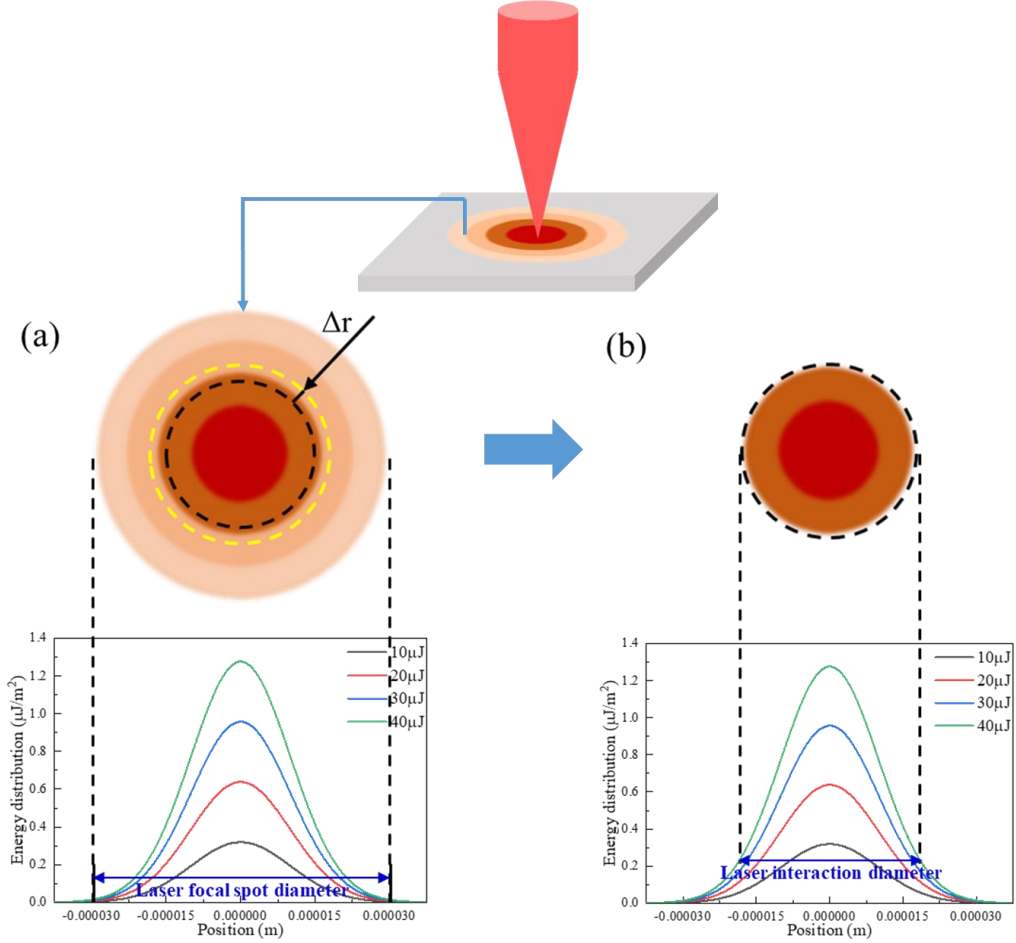

*Fig. S1. Gaussian distribution of pulse laser energy and the actual interaction area*

Considering the extremely short interaction period between the pulse laser and the liquid, we assumed an initial bubble existed inside the coated liquid layer after the laser interaction with the liquid, and such an initial bubble has the same size as the laser interaction diameter $r_T$ (30, 31). Without considering the effect of pressure change, the latent heat $E_L$ and the sensible heat $E_S$ can be calculated by Eqs. (4) and (5),

$$E_L = \frac{4}{3}\rho_l \pi r_T^3 h_{fg} \tag{4}$$

$$E_S = \frac{4}{3}\rho_l \pi r_T^3 c_p (T_i - T_e) \tag{5}$$

where $\rho_l$ is the density of liquid layer, $h_{fg}$ is the latent heat, $c_p$ is the specific heat capacity, $T_i$ and $T_e$ are the initial temperature of the initial bubble and the environmental temperature, respectively.

The sum of latent heat $E_L$ and sensible heat $E_S$ should equal to $E_a$, the total absorbed energy by the liquid layer,

$$E_a = E_L + E_S \tag{6}$$

In addition, the pressure inside the initial bubble $P_i$ can be calculated by Eq. (7),

$$P_i = \frac{\rho_l}{\rho_v} P_e \tag{7}$$

where $\rho_v$ is the density of vapor under the initial temperature, and $P_e$ is the atmosphere pressure.

**CFD Modeling - governing equations**

The Rayleigh bubble dynamics model (*38*) has been widely applied to study the response of surrounding incompressible flow to the expansion of a single spherical bubble. The governing equation for the bubble expansion within liquid can be described as follows,

$$\rho_l R \frac{d^2 R}{dt^2} + \frac{3}{2} \rho_l \left(\frac{dR}{dt}\right)^2 = P_i(t) - P_\infty(t) - \frac{2\delta}{R} - \frac{4\eta}{R}\frac{dR}{dt} \tag{8}$$

where $R$ is the bubble radius, $P_i(t)$ is the pressure inside the bubble, $P_\infty(t)$ is the pressure the hydrogel flow at the infinite distance from the bubble, $\delta$ is the surface tension and $\eta$ is the coefficient of viscosity.

Because the growth and development of bubble and jet flow is a multiphase process, the Volume of Friction (VOF) model in ANSYS Fluent was employed to track the liquid-gas interface. Considering the short interaction period, the phase change between vapor and liquid was ignored in the current model. The governing equations are shown as follows,

Energy equation

$$\rho c_p (\frac{\partial T}{\partial t} + \nabla \cdot \vec{v}T) = k_{eff}\nabla^2 T + \nabla \cdot p \tag{9}$$

Momentum equation

$$\frac{\partial}{\partial t}(\rho \vec{v}) + \nabla \cdot (\rho \vec{v}\vec{v}) = -\nabla p + \mu \nabla^2 \vec{v} + \rho \vec{g} + \vec{F} \tag{10}$$

Continuity equation

$$\frac{\partial \rho}{\partial t} + \nabla \cdot \rho\vec{v} = 0 \tag{11}$$

VOF model equation

$$\frac{1}{\rho_q}[\frac{\partial}{\partial t}(\alpha_q \rho_q) + \nabla \cdot (\alpha_q \rho_q \vec{v}_q)] = 0 \qquad (12)$$

where $\rho$ is the density of mixture, $P$ is the pressure, $k_{eff}$ is the effective conductivity, $c_p$ is the heat capacity, $\mu$ is the dynamic viscosity, $\alpha$ is the volume fraction.

**Modeling – boundary conditions and properties**

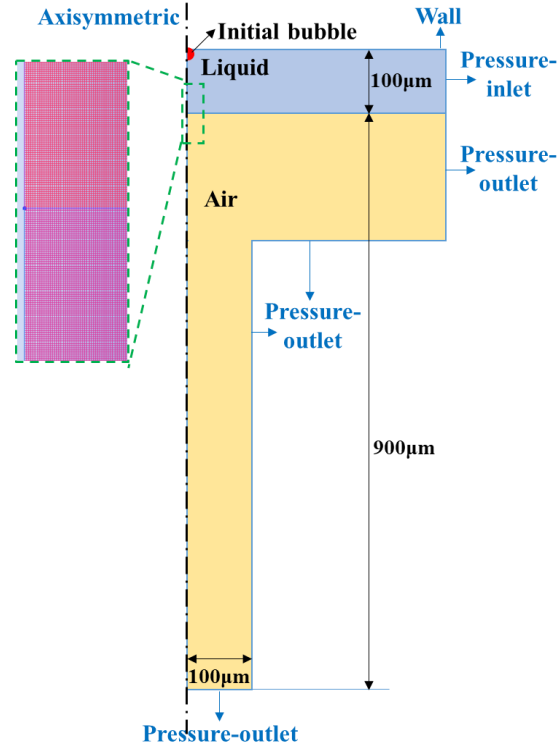

*Fig. S2. Geometry of computational domain with boundary conditions and configuration of mesh*

The dimensions of the computational domain, boundary conditions and mesh configuration were shown in Fig. S2. The computational domain includes the ribbon, the liquid layer and air. Only half of the model was meshed and simulated because of its axisymmetric geometry. Structured meshes were used in this study, and the mesh near all the boundaries was refined. Because the computational domain was only part of the ribbon, the right side of liquid was defined as the pressure inlet while the right side of air zone was defined as the pressure outlet. Besides the axisymmetric boundary condition at the axis, other boundaries were all defined as "wall". The parameters of initial bubbles were set before simulation started.

The physical properties of liquid layer were shown in Table S1. 65%-glycerol and deionized water were utilized in the simulation.

*Table S1. Physical property parameters of liquid layer*

| Properties | Density $\rho$ (kg/m³) | Heat capacity $c_p$ (kJ/(kg·°C)) | Latent heat ΔH (kJ/kg) | Viscosity $\mu$ (kg/(m·s)) | Surface tension $\sigma$ (N/m) |
|---|---|---|---|---|---|
| 65%-glycerol | 1169.1 | 3.030 | 1426.1 | 0.0177 | 0.068 |
| Deionized water | 998.2 | 4.182 | 2257.2 | 0.001003 | 0.0728 |

**Modeling – mesh independent study**

The grid independent study was carried out to a reasonable mesh number by considering the balance of computational load and numerical accuracy. Fig. S3 shows the comparison of maximum liquid velocity at 1μs among various six different cases. Apparently, the case with 580000 meshes is the most appropriate one with reasonable computational load and great numerical accuracy, since its maximum velocity change is smaller than 0.5% when the number of grid cells further increases. Therefore, the grid number of 5800000 is sufficient, and similar grid sizes were used in this study for all other CFD cases.

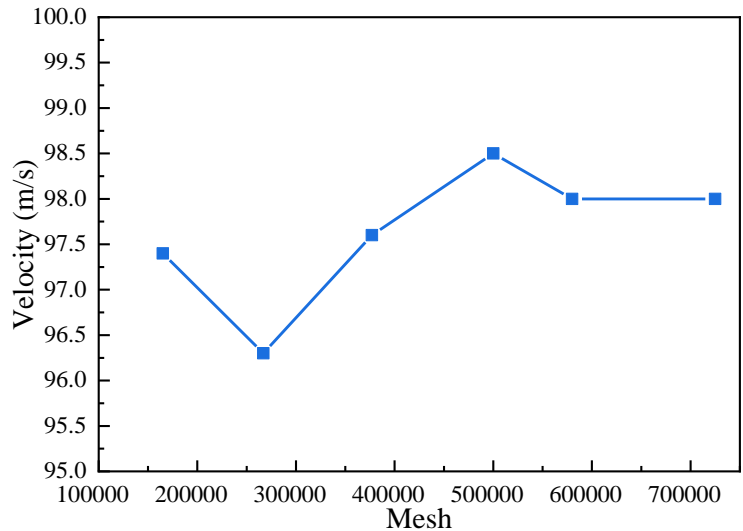

*Fig. S3. Grid dependence analysis*

**Prediction of droplet size – Static equilibrium equation**

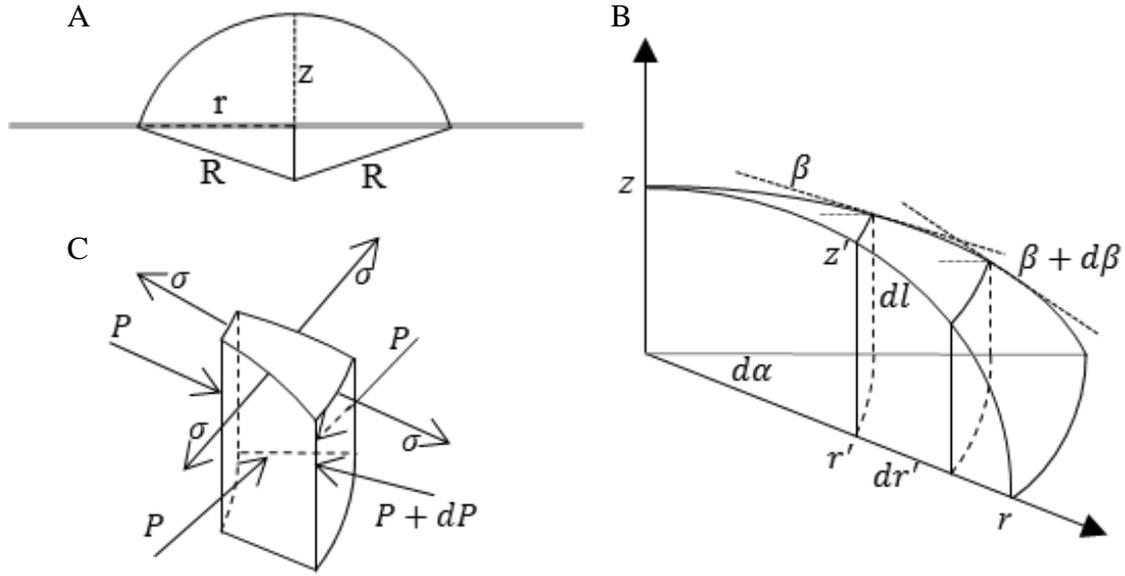

*Fig. S4. Geometry of part of sphere shape droplet and the force analysis*

Considering the droplet as part of the sphere shape, a geometric model was established to show a differential control volume, where $r$ is the radius of the droplet, $z$ is the height of the droplet, and $R$ is the radius of the sphere, as shown in Fig. S4A.

The governing equation of the static equilibrium model for the differential control volume (Fig. S4B and C) is shown as follows,

$$(-\sigma r' \sin\beta d\alpha d\beta + \sigma \cos\beta d\alpha dr') - 2\sigma dl \sin\frac{d\alpha}{2} + (-Pr'd\alpha dz' - Pz'd\alpha dr' - r'z'd\alpha dP) + 2Pz'dr'\sin\frac{d\alpha}{2} = 0 \quad (13)$$

If integrating on both sides of Eq. (13), we can obtain the governing equation of the static equilibrium model as follows,

$$-2\pi\sigma(l - r\cos\theta) + 2\pi\sigma\left[R\arcsin\left(\frac{r}{R}\right) + \frac{r(z^2 - r^2)}{z^2 + r^2}\right] + \rho g\pi\left[R^3 \arcsin\left(\frac{r}{R}\right) - \frac{1}{4}\frac{r^5}{z^2} + \frac{1}{4}z^2 r - \frac{2}{3}r^3\right] = 0 \quad (14)$$